\documentstyle[12pt]{article}

\def\apj{Astrophys. J.}
\def\apjl{Astrophys. J.}
\def\apjs{Astrophys. J. Suppl.}
\def\mnras{Mon. Not. R. Astron. Soc.}
\def\aap{Astron. Astrophys.}
\def\araa{Annu. Rev. Astron. Astrophys.}
\def\nat{Nature}

\def\physrep{Phys. Rep.}

\def\Mpc{\, h^{-1} \, {\rm Mpc}}

\def\spose#1{\hbox to 0pt{#1\hss}}
\def\lta{\mathrel{\spose{\lower 3pt\hbox{$\mathchar"218$}}
     \raise 2.0pt\hbox{$\mathchar"13C$}}}
\def\gta{\mathrel{\spose{\lower 3pt\hbox{$\mathchar"218$}}
     \raise 2.0pt\hbox{$\mathchar"13E$}}}

\begin{document}
%\baselineskip=24pt
%\title{HOW SMOOTH IS THE UNIVERSE ON LARGE SCALES ? } 
\title{THE LARGE-SCALE SMOOTHNESS OF THE UNIVERSE} 

\author{Kelvin K. S. Wu, Ofer Lahav \&~Martin J. Rees\\
{\tenrm Institute of Astronomy, Madingley Road, Cambridge CB3 0HA, UK}}

\date{\today}

\maketitle

%\section*{Abstract}
\section*{}
{\bf New measurements of galaxy clustering and background radiations
  provide improved constraints on the isotropy and homogeneity of the
  Universe on large scales.  In particular, the angular distribution
  of radio sources and the X-Ray Background probe density fluctuations
  on scales intermediate between those explored by galaxy surveys and
  the Cosmic Microwave Background experiments.  On a scale of $\sim
  100 \Mpc$ the rms density fluctuations are at the level of $\sim 10
  \%$ and on scales larger than $300 \Mpc$ the distribution of both
  mass and luminous sources safely satisfies the `Cosmological
  Principle' of isotropy and homogeneity.  The transition with scale
  from clumpiness to homogeneity can be phrased in terms of the
  fractal dimension of the galaxy and mass distributions.}

\bigskip

%\section{Introduction}

\section*{}
Most cosmologists believe that on the very large scales
the Universe closely obeys the equations of General Relativity for an isotropic
and homogeneous system---this is the 
so-called `Cosmological Principle'\cite{peebl80,peebl93}.
However, on scales much smaller than the horizon the distribution of luminous
matter (i.e. galaxies) is clumpy (see Figure 1).  
This is commonly attributed to gravity, which amplifies the
tiny initial density contrasts in the mass distribution as the universe
expands.  The Cosmological Principle was first adopted when observational
cosmology was in its infancy; it was then little more than a
conjecture. Observations could not then probe to significant redshifts, the
`dark matter' problem was not well-established and the Cosmic Microwave
Background (CMB) and X-Ray Background (XRB) were still unknown.  
If the  Cosmological Principle turned out to be invalid 
then the consequences to our understanding of cosmology would be dramatic, 
for example the conventional way of interpreting the age of the universe, 
its geometry and matter content would have to be revised. 
Therefore it is 
important to revisit this underlying assumption in the light of new
galaxy surveys and measurements of the background radiations.
The question of whether the universe 
is isotropic and homogeneous on large scales
can also be  phrased in terms of the fractal structure of the 
universe\cite{mande83,mande98, cps88,pms97,smp98, davis97}.
A fractal is a geometric shape that is not homogeneous, 
yet preserves the property that that each part is a reduced-scale
version of the whole.
If the matter in the universe were actually 
distributed like a pure fractal on all scales then the 
Cosmological Principle 
would be invalid, and the standard model in trouble.

Past attempts to address the above questions have confronted  
two gaps: 
(i) Most of
the gravitating material is dark, 
and it is still unclear how to relate
the distributions of light and mass, in particular how to match the clustering
of galaxies with the CMB anisotropies, which tell us about the mass
fluctuations.
(ii) 
Little is known about fluctuations on intermediate scales between those of
local galaxy surveys ($\sim 100 \Mpc $, where $h$ is the Hubble constant in
units of 100 km/sec/Mpc) and the scales probed by the Cosmic Background
Explorer (COBE) satellite ($\sim 1000 \Mpc $).

In this review we summarize what can be learnt from various recent observations
and techniques for studying clustering on large scales.  We consider in
particular probes such as radio sources and the XRB, at median redshift ${\bar
z \sim 1} $, which seem to cover effectively these intermediate scales.  CMB
experiments on smaller angular scales than those probed by COBE are also
helping to bridge the gap, and future big redshift surveys of more than one
million galaxies will probe to median redshift ${\bar z} \sim 0.1$
(which roughly corresponds to a comoving distance of $\sim 300 \Mpc$).
Current data already strongly constrain any non-uniformities in the 
galaxy distribution (as well as the overall mass distribution) 
on scales $\gta 300 \Mpc$.

Any quantitative discussion of the large scale structure in the universe
actually depends on the unknown cosmological parameters 
(defined only for a homogeneous 
and isotropic universe):
the density parameter
$\Omega$, the cosmological constant $\Lambda$ and the Hubble constant $H_0$.
For simplicity we present the observational results interpreted for the
Einstein-de Sitter model ($\Omega=1$ and $\Lambda=0$), but the main conclusions
are not altered for other models.  

\section*{Local galaxies are strongly clustered}

The clumpiness of matter in the Universe was initially studied
by measuring the clustering of bright galaxies\cite{peebl80,peebl93}.
Figure 1\nocite{mesl90} shows the distribution of 2 million optically selected
galaxies\cite{mesl90} projected on the sky. 
The distribution is evidently not uniform: 
galaxies are arranged in clusters and  `filaments', and avoid
certain regions termed `voids'. 
Figure 2\nocite{shect96} shows data from the largest 
redshift survey to date, Las Campanas\cite{shect96}, which
illustrates (insofar as the redshift of each galaxy indicates its
distance) the three-dimensional clustering of galaxies.
Although clustering is seen on scales of tens of Mpc's, 
on larger scales the distribution seems more homogeneous.

More quantitatively, 
it is  well established that 
the probability of finding a galaxy 
$\sim 5 \Mpc$ away from  another galaxy is twice the probability 
expected in a uniform distribution (see Box 1).  
The clustering of optical galaxies\cite{be93,be94} is illustrated in
Figure~3\nocite{be93,be94,smoot92,benne96,hanco97,gs98,bllw98,lpt97,treye98}.  The plot shows the
fluctuations $\langle({ {\delta \rho} \over \rho })^2 \rangle$ 
as a
function of characteristic length scale $\lambda$. 
The solid and dashed lines correspond to two
variants of the
Cold Dark Matter (CDM) model for mass density fluctuations\cite{bbks86},
which is widely used as a `template' for comparison with data.
We see that the fluctuations drop monotonically with scale
(although not as a pure power-law). 
On a  scale of $\lambda \sim 100 \Mpc$ the rms fluctuation is
only $10 \%$. This is the key evidence that on  larger and larger scales
the fluctuations become negligible.

As mentioned above 
it is most unlikely that luminous galaxies trace perfectly the mass
distribution. 
Galaxies can only form in dense regions, 
and their formation may be affected 
by other physical conditions and local  environment.
Hence the clustering of galaxies is likely to be `biased' 
relative to the mass fluctuations (see Box 1).
Indeed the galaxy
distribution could in principle display conspicuous features on very
large scales even if the mass did not --- for instance, a long cosmic
string could `seed' galaxy formation in its wake.  So the galaxies
could be arrayed in a fractal structure, even if the mass distribution
is non-fractal.  
It is important therefore to understand the biasing in order to match 
the fluctuations in galaxies to the fluctuations in mass 
in Figure 3.

Other (biased) probes at large distances are 
clusters of galaxies, as selected optically by Abell\cite{abell58}
or by X-ray surveys\cite{ebeli96}.
These surveys typically probe out to redshift $z\sim 0.1$.
Several studies\cite{scara92,scara98}
suggest that on  scales of $\sim 600 \Mpc$  the distribution of Abell clusters 
is homogeneous.

A more controversial result on the distribution of galaxies suggests a
`characteristic scale' of clustering of $\sim 128
\Mpc$\cite{beks90,landy96}.
It is not clear yet if this feature is real, or
just due to small number statistic or survey geometry\cite{kp91}.
Recently Einasto et al.\cite{einas97} suggested that the distribution of Abell
clusters is a quasiregular three-dimensional network of superclusters and
voids, with regions of high density separated by about $120 \Mpc$.  The reality
of such a `periodicity' in galaxy clustering will soon be revisited by 
two new large redshift surveys.  
The American-Japanese Sloan Digital Sky Survey (SDSS)
will yield redshifts for about 1 million galaxies and the Anglo-Australian `2
degree Field' (2dF) survey, 
will produce redshifts for 250,000 galaxies (both with
median redshift of ${\bar z} \sim 0.1$).  These big galaxy surveys will give
good statistics on scales larger than $\sim 100 \Mpc$.

\section*{Tracers at high redshift}

We need to observe beyond $z=0.1$ in order to
sample a big enough volume to probe clustering on scales above $300
\Mpc$ and to fill the gap between scales probed by galaxy
surveys and the scales probed by COBE. However this then introduces
the extra complication that we cannot interpret the data without
taking account of how the clustering evolves with time, and also
possible cosmic evolutionary effects in the brightness of
objects. Here we discuss the
X-Ray Background (XRB) and radio sources as probes with median
redshift $ {\bar z} \sim 1$.  Other possible high-redshift tracers are
quasars and clusters of galaxies.

The XRB and radio sources are tracers of galaxies (or at least of that
subset which is active).  We see from Figure 3 that the limits they
set to large scale inhomogeneities ($>100 \Mpc$) are less stringent
than those implied by CMB measurements, but  they provide 
independent constraints, since they sample `luminous' objects 
rather than the total mass.

\subsection*{Radio galaxies} 
Radio sources in surveys have typical median redshift
${\bar z} \sim 1$, and hence are useful probes of clustering at high
redshift. 
Earlier studies\cite{webst76} 
claimed that  the distribution of radio sources supports the 
`Cosmological Principle'.
However, the redshift distribution of radio sources is now 
better understood
and it is 
clear that the wide range in intrinsic luminosities of radio sources
would dilute any clustering when projected on the sky.  
Recent analyses  of
new deep radio surveys\cite{condo98}
suggest that radio sources are clustered at least as strongly
as local optical
galaxies\cite{sp89,bw95,kbk95,sicot95,lwl97,chbgw96,mmlw98}.
Nevertheless, on very large scales the distribution of radio sources
seems nearly isotropic. 
Comparison of the measured quadrupole in a radio sample 
to the theoretically predicted ones\cite{bllw98}
offers a crude estimate of the fluctuations on scales $ \lambda \sim 600
h^{-1}$ Mpc.  The derived amplitudes are shown in Figure 3 for the two 
assumed CDM models.
Given the problems of catalogue matching and shot-noise, these points should be
interpreted as significant `upper limits', not as detections.  

\subsection*{The X-Ray Background} 
Although discovered in 1962, the origin of the
X-ray Background (XRB) is still controversial, but its sources, whatever they
turn out to be, are likely to be at high redshift\cite{boldt87,fb92}.

The XRB is a unique probe of fluctuations on intermediate scales
between those of local galaxy surveys and
COBE\cite{rees80,shafe83,bct97,lpt97,bfc97,jahod98,bough98}, although
the interpretation of the results depends somewhat on the nature of
the X-ray sources and their evolution.  The rms dipole and higher
moments of spherical harmonics can be predicted\cite{lpt97} in the
framework of gravitational instability 
and assumptions on the distribution of the X-ray sources with redshift. 
By comparing
the predicted multipoles to those observed by HEAO1\cite{treye98}
it is possible to estimate the amplitude of fluctuations for an
assumed shape of the density fluctuations 
(e.g. CDM model).  
Figure 3 shows the amplitude of fluctuations derived at the 
effective scale $\lambda \sim 600 \Mpc$ probed by the XRB. 
Assuming a specific epoch-dependent biasing scheme\cite{fry96}
(see Box 1)   
and a range of models of evolution of X-ray sources and clustering,
the present-epoch density fluctuations of 
of the X-ray sources is found 
to be no more than twice
the amplitude of fluctuations in mass\cite{treye98}.
Below we shall use this estimate to derive a fractal
dimension of the universe on large scales, which turns out to be
indistinguishable from the one for a homogeneous universe.

\subsection*{Quasars, high redshift galaxies, and Lyman-$\alpha$ clouds}
                              
Until the mid-1990s, quasars -- hyperactive galactic nuclei -- were
the only objects luminous enough to be identified in substantial
numbers at redshifts $z> 2$.  It is still unclear how their clustering
evolves with redshift\cite{sb94}, but there is no evidence that they
are any more clustered than extragalactic radio sources (which are a
related population).  The advent of 10-metre-class telescopes now
allows the detection of galaxies out to equally large redshifts.
Already, several hundred galaxies with $z = 3$ have been detected, and
they display about the same level of clustering as nearby
galaxies\cite{steidel97}.  Since gravitational effects enhance density
contrasts during cosmic expansion, one might at first sight have
expected weaker clustering of galaxies at earlier times. However, the
luminous galaxies that have already formed at the epoch corresponding
to $z = 3$ belong to an exceptional subset associated with
unusually high density peaks, which display enhanced clustering (see Box
1). When this is taken into account, the data are compatible with CDM
models normalised to match the degree of clustering at low
redshifts\cite{steidel97}.  Larger samples of high-$z$ galaxies will
soon provide direct evidence on the clustering of early galaxies on
scales up to and beyond $100 \Mpc$; comparison with the lower-redshift
samples will then elucidate exactly how the clustering evolves. Such
data will perhaps reveal larger-scale clustering at high redshifts,
but the amplitude of this is already constrained by the radio sources
and XRB data.  The rich absorption-line spectra of high redshift
quasars offer a probe for the distribution of intervening material.
In particular, the `Lyman-$\alpha$ forest' in quasar spectra reveals the
distribution of gas clouds (probably intimately related to low-mass
protogalaxies) along the line of sight.  The absorption features in
each spectrum are smoothly distributed in redshift.  There are no
large `clearings' in the Lyman-$\alpha$ forest: indeed the relevant
clouds seem even more smoothly distributed than galaxies
\cite{wb91}. Moreover, the spectra of different quasars indicate that
the clouds have the same properties along all lines of sight. Relating
the spacings of these clouds to the overall mass distribution (or even
to the galaxy distribution) is not straightforward.  However, if these
distributions manifested scale-free fractal-like structure it would be
remarkable if this structure were not plainly apparent in quasar
absorption spectra\cite{davis97}.

\section*{Direct probes of mass distribution} 

As already discussed, the luminous objects selected by surveys may not
trace the total mass. There are however at least three
independent probes of inhomogeneities in the  gravitational
field induced by the total mass fluctuations:
lensing, the CMB, and peculiar velocities. 
Gravitational lensing-- the distortion of distant
galaxy images  by intervening potential
wells --can only  constrain the mass fluctuations
on scales smaller than $20 \Mpc$ or so\cite{villu96,kaiser96}.
%KAISER (astroph/9610120)
Here we focus on the other two probes of mass fluctuations, 
which show good consistency with the picture that fluctuations 
are significant on small scales (as deduced from peculiar velocities) 
but are tiny 
on the very large scales (as inferred from the CMB).

\subsection*{The CMB}
The CMB is well described by a black-body radiation spectrum at a temperature
of $2.73^\circ K$, hence providing crucial evidence for the hot Big Bang model.
This sea of radiation is highly isotropic, the major anisotropy being due to
the motion of our Galaxy (the Milky Way) at 600 km/sec relative to the CMB.
This motion is remarkably reconstructed in both amplitude and 
direction by summing up the forces due to  masses 
represented by galaxies\cite{Strauss92,wlf96} 
at distances nearer than $\sim 100 \Mpc$.
The dipole anisotropy in the distribution of nearby 
supernovae also indicates 
that most of the Galaxy's  motion arises
from local inhomogeneities\cite{riess97}. 
The agreement between the CMB dipole and the dipole anisotropy of 
relatively nearby galaxies 
is a good argument in favour of large scale homogeneity.
In an arbitrarily lumpy universe this would be a coincidence.
A given overdensity 
${\delta \rho} $ on a scale $\lambda$ produces a peculiar velocity
proportional to $\lambda$ (in the linear regime), so the absence of very large
peculiar velocities is in itself evidence that ${\delta
\rho}$ decreases as steeply as $\propto 1/\lambda$.

Apart from the dipole anisotropy, the other main anisotropies in the
CMB radiation are imprints of the potential wells at the last
scattering surface and their influence on the plasma motions at this
epoch.  As the photons climbed out of different gravitational
potential wells they experienced gravitational redshift.  On angular
scales larger than $1^\circ$, separate regions on the sky were not
causally connected.  In 1992 the COBE satellite detected
fluctuations in the CMB, at the level of $10^{-5}$ on scales of
$10^\circ$; which corresponds to a present-epoch length-scale of $\sim
1000 \Mpc$ (see Figure 3).
These tiny CMB fluctuations are attributed to `metric' or `curvature'
fluctuations\cite{sw67} of this order in a universe which has a
Friedmann-Robertson-Walker (FRW) metric of space-time (corresponding to
a homogeneous and isotropic universe).  
Moreover, the concept of `inflation' has suggested a
physical reason why the universe may end up close to FRW metric but
with fluctuations in the metric that are equal on all scales.  
Further support for this
hypothesis comes from studies of how the present-day clustering of
galaxies might have evolved via gravitational instability: the
relevant fluctuations here are on much smaller scales than those
probed by COBE but the required amplitude is again of order of
$10^{-5}$.  
The metric fluctuations on larger scales would 
not have evolved into bound structures: the associated density contrast
would still be small, but for constant metric fluctuations
it would have an amplitude inversely proportional 
to the square of the length-scale.

On scales smaller than 1 deg 
causal physical processes took place.
The plasma oscillated acoustically in response to the 
fluctuations in the potential wells of the dark matter.
This interaction between plasma and gravity translates to peaks
in the angular power-spectrum of the temperature
fluctuations\cite{wss94,ssw95,hss97,btw97}.
The angular scales of  these peaks correspond to few hundreds of
$\Mpc$ today, hence can be compared with other  probes 
of the fluctuations such as galaxies, radio sources and the XRB (see Figure 3).
The existing and 
new experiments (e.g. Tenerife, CAT, Saskatoon, Planck, MAP, VSA) will map 
the   fluctuations on these scales.
This is important because the position and height of the peaks can be used to 
determine the cosmological parameters with very high
precision\cite{hss97,btw97,gs98,whllr98}.
Also,
these smaller-scale fluctuations are the precursors of those which
have developed into clusters and superclusters by the present time.   
Indeed, the imprint of 
fluctuations represented by the CMB angular `peaks',
even though at the level of few percent, 
might show up 
in the deep galaxy maps soon to be produced by 
the Sloan and 2dF surveys.

One may question if large amplitude inhomogeneities along the line of
sight can wash out large intrinsic fluctuations in the CMB and make it
look very smooth.  The general effect would however be merely to
distort the the angular distribution of the fluctuations rather than
to homogenize the temperature map\cite{slejak97}.

\subsection*{Peculiar velocities} 
Peculiar velocities (like the 600 km/sec motion 
of the Local Group described above) 
are deviations from the recession velocity 
that would be expected due to the smooth 
the expansion of the Universe. Their measurement 
in other galaxies requires
redshift-independent distances, derived by observing the
apparent brightness or size of a galaxy and relating this to some
parameter, e.g. internal stellar velocities, that is known to be a
measure of its intrinsic luminosity.  The gross features of the local
peculiar velocity field inferred in this way 
correlate well with overdensities in galaxy
distribution, e.g. the Virgo cluster and the Great
Attractor\cite{lynde88,dekel94,sw95}, although in some regions the
agreement is not perfect, perhaps due to systematic measurement
errors.

Unfortunately, the distance measurement errors increase with
distance, so the observed peculiar velocity field\cite{kd97} can only
probe scales $\lambda < 20 \Mpc$.  Lauer and Postman\cite{lp94} claimed that a
sample of Abell clusters out to $150 \Mpc$ is moving at $\sim 700 $ km/sec with
respect to the CMB, suggesting that the CMB dipole (caused by such relative
motion) is generated largely by mass concentrations beyond $\sim 100 \Mpc$, but
most other studies suggest bulk flows on smaller scales.

\section*{Fractal vs.\ homogeneity on large scales}

A fractal is a distribution or shape that is not homogeneous, but
possesses the property that each part of it is a version of the whole
reduced in scale. It other words it `looks' the same on all scales:
from `afar', or close up for an enlarged view of some portion.
Fractals abound in nature and are fundamental in
physics\cite{mande83,mande98}, for they accompany the basic idea of
the scaling of physical laws. For example, the coastline of a small
peninsula drawn on paper could look equally valid for a large
continent; and fractals have long been studied in solid state physics.
The clustering of galaxies (see Fig. 1) lends itself to a fractal
description since the clumpiness prevails over a wide range of scales.
In the language of fractals (see Box 2), fractal dimensions are used
to characterise the degree of clustering.  Fractal dimensions
generalise our intuitive concepts of dimension and can take
any positive value less than 3. 
On scales below $\sim 10\Mpc$ galaxies are distributed
with the correlation dimension\cite{peebl80}  $D_2= 1.2-2.2$
(see also Box 2 and Table~1).  
This is well below the homogeneous value of 3.

The other side of the coin is that the mass distribution has to
approach homogeneity on large scales in order for the FRW metric,
the standard model of space-time, to
hold.  To reconcile the two models one supposes that the fractal model
crosses over gradually to homogeneity, i.e.  $D_2=3$ (see Figure 4).
Although luminous matter
does not necessarily trace mass, the detection of this large-scale
homogeneity is naturally a very important quest. If the galaxy
distribution were a fractal on very large scales, it may imply seeding
of galaxy formation by topological defects (e.g. strings) uncorrelated 
with the large scale mass distribution,
or it may even have important implications for the application of the
FRW metric, unlikely though this may be given the many successes of
the `standard cosmology'.

In recent years Pietronero and coworkers\cite{cps88,pms97,smp98} have
strongly advocated that the scale of homogeneity has not been detected
even in the deepest redshift surveys. Most analyses of density
fluctuations  had {\em assumed} large-scale homogeneity but these authors
applied methods that made no such
assumptions and argued that the fractal behaviour extended to the
largest scales probed ($\sim 1000\Mpc$), with $D_2\approx 2$.
However, Cold Dark Matter models of density fluctuations (which fit
reasonably well the available observational data) predict that at
scales above $\sim 10\Mpc$ one should begin to detect values of $D_2$
greater than 2, with $D_2 \approx 3$ on scales larger than $\sim
100 \Mpc$ (see Box 2 and Figure 4), in conflict with Pietronero's
claim.

Several authors have therefore made further analyses of galaxy
distributions using fractal
algorithms\cite{gicgh91,mc94,ls91,borga95,scara98,cbcm98,mpmg98}. All
of them obtained results which were consistent with standard models of
density fluctuations and all appeared to detect an approach to
homogeneity on the largest scales analysed by them. We
list some results for $D_2$ in Table~1. Although the statistics are
still poor, one can already see the steady increase towards
$D_2=3$. 
In particular, the results do not support a constant $D_2$
for all scales, and the latest results by Scaramella et
al.\cite{scara98} provide the closest {\em fractal} measurement yet to
homogeneity.  Scaramella et al.\ pointed out the importance of
appropriate corrections to the observed flux
from high-redshift galaxies to account for redshifting of their
spectrum across the observer's waveband. 
These spectral band corrections are crucial to the interpretation of
high-redshift photometry and should not be left out, approximate
though they may be.
On smaller scales, no other group has
substantiated the value $D_2\approx 2$ on all scales larger than
$10\Mpc$. Lemson \& Sanders\cite{ls91}, 
who did obtain this value on scales below
$30\Mpc$ (Table~1), also detected a crossover to
homogeneity for larger scales.

The debate has therefore brought up technical issues which have been
useful. For example, it was correctly pointed out\cite{cps88} that
if a survey is too small, then one cannot define the mean density (for
the galaxies do form a fractal on small scales) and hence related
tools such as correlation functions can be misleading. On the other
hand, the fractal proponents have not helped their cause by using
highly incomplete and inhomogeneous samples.
%(such as the LEDA database).
Detailed technical arguments for\cite{smp98} and
against\cite{davis97, guzzo97, calzetti91} fractals on large scales have been given. 
The continued application of fractal algorithms
(as opposed to traditional methods which assume homogeneity) to larger
and deeper surveys should definitively resolve the matter.

Alternative arguments against the large-scale fractal model have been
given by Peebles\cite{peebl93}. He pointed out that the variation of
both the number counts and the angular correlation 
function of galaxies with apparent luminosity point strongly against a pure
fractal universe.  Furthermore, since properties of a pure fractal
are independent of scale, the projected galaxy distribution in shells
of increasing size should look the same.
This is strongly in conflict with the
observations\cite{davis97}, which show decreasing clumpiness in larger
shells. However, we note that visual impression alone cannot indicate
the closeness to isotropy.

Direct estimates of $D_2$ are not possible for much larger scales, but
we can calculate values of $D_2$ at the scales probed by the XRB and
CMB by using CDM models normalised with the XRB and CMB as described
above.  The resulting values are extremely close to 3 and are given in
the lower part of Table~1. (They are even tighter than the constraint
$3-D_2\leq 0.001$ obtained by Peebles\cite{peebl93} from the XRB using
a different argument.)  We consider the agreement of XRB and CMB
fluctuations with the popular CDM models {\em within the framework} of
a homogeneous universe to argue strongly against a pure fractal galaxy
or mass distribution.  (We remind the reader that the XRB
traces galaxies, while the CMB traces mass.) Can we go further?
Isotropy does not imply homogeneity, but the near-isotropy of the CMB
can be combined with the Copernican principle that we are not in a
preferred position.  All observers would then  measure the same
near-isotropy, and an important result 
has been proven that 
the universe must then be very well approximated by 
the FRW metric\cite{sme95,mes96}.

While we reject the pure fractal model in this review, the performance
of CDM-like models of fluctuations on large scales have yet to be
tested without assuming homogeneity {\it a priori}. On scales below,
say, $30\Mpc$, the fractal nature of clustering implies that one has
to exercise caution when using statistical methods which assume
homogeneity.  As a final note, we emphasize that we only considered
one `alternative' here, which is the pure fractal model where $D_2$ is a
constant on all scales.

\section*{Discussion}

To study galaxy evolution and the validity of the FRW metric on large
scales it is important to explore density fluctuations at higher
redshift. The examples of the X-ray Background and radio sources are
encouraging, but redshift information is required to constrain the
growth of cosmic structure with time.  Our main conclusion is that,
while the galaxy distribution can be described as a fractal on scales
smaller than $20 \Mpc$, there is strong evidence from present
data on the XRB and the CMB that 
homogeneity and isotropy approximately prevails
 on scales larger than $300 \Mpc$.
However, the scale of `cross-over' to homogeneity is not well
determined yet.  We emphasize that it is difficult to `prove' the
Cosmological Principle.  However, the recent observational tests
described in this review, in combination, offer extremely strong
support for this crucial hypothesis.

We have derived constraints on large-scale ($>100$ Mpc) irregularities
in the distribution of galaxies and other luminous objects. 
These could have turned out to be much less
homogeneous than the overall mass distribution. On the largest scales
($\sim 1000\Mpc$) possible inhomogeneities in the matter distribution
are strongly constrained by the smallness of the CMB fluctuations --
it is this evidence which tells us, most convincingly, how remarkably
accurately the Friedmann-Robertson-Walker models seem to fit our
universe.  Of course, we cannot formally exclude a pre-Copernican
model universe, such that the isotropy around us is atypical of what
would be measured by hypothetical observers elsewhere\cite{ellis84}.
But, leaving such aside, there is a well-defined sense in which our
universe is homogeneous on the largest accessible scales; neither its
mass distribution, nor that of the galaxies
resembles a pure fractal.  Cosmological
parameters such as $\Omega$ therefore have a well-defined meaning ---
indeed these considerations tell us over what volume we need to
average in order to determine them with any specified level of precision.

%\nocite{*}

\section*{Acknowledgments}
We thank Audra Baleisis, Eric Gawiser, Luigi Guzzo and Marie Treyer
for helpful discussions.

\section*{Box 1: Quantitative measures of galaxy clustering} 

One popular measure of galaxy clustering is the two-point correlation 
function\cite{peebl80},
defined as the excess
probability, relative to a random distribution,  of finding a galaxy 
at a distance $r$ from another galaxy. 
It is now well established that on scales smaller than $\sim 10 \Mpc$ 
it has roughly the form
\begin{equation}
\xi(r) = ({ r \over r_0 })^{-\gamma}\;.
\end{equation}
For optically selected galaxies $\gamma =1.8$, $r_0 \approx 5 \Mpc$,
while for galaxies observed in the infrared with the IRAS satellite, which
include spiral galaxies but under-represent ellipticals,
$r_0 \approx 4 \Mpc$ with a somewhat shallower slope\cite{srl92}.
The clustering of galaxy clusters (as selected by Abell\cite{abell58}
or by X-ray surveys) 
obeys a similar law\cite{bahca88}
but with a much stronger clustering amplitude, $r_0 \approx 15-20 \Mpc$.
The Fourier transform of the correlation function $\xi(r)$ 
is the power-spectrum $P(k)$ 
(where $k$ is the wavenumber),
which corresponds to the square of Fourier
coefficients of the fluctuations.
The rms fluctuations (see Figure 3) 
can be written as 
 $\langle({ {\delta \rho}
  \over \rho })^2 \rangle \propto k^3 P(k)$.

It is not likely that the fluctuations in the density 
of particular galaxy types
are exactly the same as the fluctuations in mass.
The simplest assumption, which has been widely adopted,
is that the galaxy and mass density fluctuations at any point 
${\vec {x}}$ are
related by
\begin{equation}
\delta_g(\vec{x}) = b \delta_m(\vec{x}) ,
\end{equation}
where $b$ is the `bias parameter'. Usually $b>1$ which implies that
the galaxies are more clustered than the mass distribution. By
modelling galaxies as peaks of the underlying mass distribution and
using an argument analogous to that which explains why the highest
ocean waves come in groups, Kaiser\cite{kaise84} showed that in the linear
approximation the correlation function of galaxies is related to the
mass correlation function by
\begin{equation}
\xi_{gg}(r) = b^2 \xi_{mm}(r), 
\end{equation}
where $r$ is the separation between galaxies or mass elements.
Note that although
eq. (3) does follow from eq. (2), it is more general and does not imply
eq. (2).  Various theoretical and observational considerations suggest that
$b\approx 1-2$.

Biasing must certainly be more complicated than eqs. (2) and (3): indeed, 
clustering is not the same for galaxies of different galaxy morphologies.
For example, elliptical galaxies are more strongly clustered than 
spiral galaxies on scales $\lta  10 h^{-1}$ Mpc\cite{dress80,lmep95,hermi96}.
The appropriate value of $b$ may depend on scale, as well as on the local
overdensity. 
Furthermore, it is not clear {\it a priori} that $\delta_g$ is
just a function of $\delta_m$. The efficiency of galaxy formation
could in principle be modulated by some large-scale environmental
effects (e.g.\ heating by early quasars, or the proximity of a cosmic
string) which are uncorrelated with $\delta_m$.
Biasing might therefore 
be non-local, non-linear, stochastic and epoch-dependent
\cite{dr87,kns97,fry96,bagla98,babul91,dl98,steidel97}.

\section*{Box 2: The fractal dimension} 

If we count, for each galaxy,
the number of galaxies within a distance $R$ from it, and call the
average number obtained $N(<R)$, then the distribution is said to be a
fractal of correlation dimension $D_2$ 
if $N(<R)\propto R^{D_2}$. Of course $D_2$
may be 3, in which case the distribution is homogeneous rather than
fractal.  In the pure fractal model this power law holds for all
scales of $R$, whereas in a hybrid model it holds for $R$ less than
some scale, above which $D_2$ increases towards 3 to accommodate the
Cosmological Principle. To allow for varying $D_2$ one often writes
\begin{equation}
 D_2 \equiv {  {d \;ln N(<R)} \over   {d \;ln R} }\;.
\end{equation}
Using the above, the fractal proponents\cite{pms97,smp98} have
estimated $D_2\approx 2$ for all scales up to $\sim 1000\Mpc$, whereas
other
groups\cite{gicgh91,mc94,ls91,borga95,scara98,cbcm98,mpmg98}
have obtained, in general, scale-dependent values as listed in
Table~1.

These measurements can be directly compared with the popular Cold Dark
Matter models of density fluctuations, which predict the increase
of $D_2$ with $R$ for the hybrid fractal model. If we now assume
homogeneity on large scales, then the mean density ${\bar n}$ and the
correlation function $\xi(r)$ can be defined, and
\begin{equation}
N(<R) =  { 4 \pi \over 3} R^3  {\bar n} 
 + 4 \pi {\bar n} \int_0^R dr r^2 \xi(r) \;,
\end{equation}
for a flat universe with $\Omega=1$.  Hence we have a direct mapping
between $\xi$ and $D_2$. If we choose a power-law form for $\xi(r)$
(eq. 1), then it follows\cite{ss85} that $D_2=3-\gamma$ if $\xi\gg 1$.
If $\xi(r)=0$ we obtain $D_2=3$. The CDM models give us the
correlation function $\xi(r)$ on scales greater than $\sim 10\Mpc$,
where we do not need to worry about non-linear gravitational effects.
The function $N(<R)$ can then be calculated from these correlations.
The predicted runs of $D_2(R)$ from three different CDM models are
given in Fig.~4. They may differ somewhat but they all show the same
qualitative behaviour: above $30\Mpc$ we should be able to measure
dimensions close to 3, not 2. Above $100\Mpc$ they become
indistinguishably close to 3. They also illustrate that it is
inappropriate to quote a single crossover scale to homogeneity, for
the crossover is gradual.  Here we have described but one statistical
fractal measure, $D_2$, out of a much larger set known as the
`multifractal spectrum', which is a useful tool for the statistical
description of redshift surveys \cite{martinez91}.

\section*{Figures and Table}

\subsection*{Figure 1:} The distribution of 2 million galaxies with magnitude
$17 \le b_j \le 20.5$ shown in an equal area projection centred on the
South Galactic pole.  The data from APM scans 
over a contiguous area of $4300$ square degrees\cite{mesl90}.
The small empty patches in the map are regions excluded
around bright stars, nearby dwarf galaxies, globular clusters and step
wedges. Although in projection, the pattern of the distribution of
galaxies is seen to be non-uniform, with clusters, filaments and
voids.

\subsection*{Figure 2:} The redshift distribution of more than 10,000 galaxies,
from the Las Campanas Redshift Survey\cite{shect96}. The plot shows
the superposition of 3 slices in the North Galactic cap, and likewise
for the South Galactic cap, plotted in redshift versus angular (RA)
coordinates. Clustering of
galaxies is seen on scales smaller than $\sim 30 \Mpc$, but on larger
scales the distribution approaches homogeneity.  Note that the diluted
density of galaxies at higher redshifts is an artifact, due to the
selection of galaxies by their apparent flux.

\subsection*{Figure 3:} 
A compilation of density fluctuations on different scales from various
observations: a galaxy survey, deep radio surveys, the X-ray
Background and Cosmic Microwave Background experiments.  The
measurements are compared with two popular Cold Dark Matter models.
The Figure shows mean-square density fluctuations $\langle({ {\delta \rho}
\over \rho })^2\rangle$. The
solid and dashed lines correspond to the standard Cold Dark Matter
power-spectrum (with shape parameter $\Gamma = 0.5$) and a
`low-density' CDM power-spectrum (with $\Gamma=0.2$), respectively.
Both models are normalized such that the rms fluctuation within $8
\Mpc$ spheres is $\sigma_{8,M}=1$. 
The open squares at small scales are estimates of the
power-spectrum from 3-dimensional  inversion of the angular APM galaxy
catalogue\cite{be93,be94}.  The elongated `boxes' at large scales
represent the COBE 4-yr\cite{smoot92,benne96,gs98} (on the right) and
Tenerife\cite{hanco97} (on the left) CMB measurements.  
The solid triangles
represent constraints from the quadrupole moment 
of the distribution of radio sources\cite{bllw98}.  
This quadrupole
measurement
probes fluctuations on scale $\lambda_*^{-1}
\sim 600 h^{-1}$ Mpc.  
The top and bottom solid triangles are upper
limits of the amplitude of the power-spectrum at $\lambda_*$, assuming
CDM power-spectra with shape parameters $\Gamma=0.2$ and $0.5$
respectively, and an Einstein-de Sitter universe.
The crosses represent constraints from the XRB HEAO1
quadrupole\cite{lpt97,treye98}.  Assuming evolution, clustering and
epoch-dependent biasing prescriptions, this XRB quadrupole measurement
probes fluctuations on scale $\lambda_*^{-1} \sim 600 h^{-1}$ Mpc,
very similar to the scale probed by the radio sources.  The top and
bottom crosses are estimates of the amplitude of the power-spectrum at
$\lambda_*$, assuming CDM power-spectra with shape parameters
$\Gamma=0.2$ and $0.5$ respectively, and an Einstein-de Sitter
universe.  The fractional error on the XRB amplitudes (due to the
shot-noise of the X-ray sources) is about 30\%.

\subsection*{Figure 4:}

The fractal correlation dimension $D_2$ versus
length scale $R$ assuming three Cold Dark Matter models of
power-spectra with shape and normalization parameters ($\Gamma =0.5$;
$\sigma_8=0.6$), ($\Gamma =0.5$; $\sigma_8=1.0$) and ($\Gamma =0.2$;
$\sigma_8=1.0$).  
Regardless of model they all exhibit the same qualitative behaviour of
increasing $D_2$ with $R$, becoming vanishingly close to 3 for $R >
100 \Mpc$.
Pietronero's pure fractal model\cite{cps88,pms97,smp98} 
corresponds to the horizontal axis $D_2=2$, 
in conflict with CDM models and the data presented in Table 1.

\newpage
\subsection*{Table 1}
Estimates of the fractal correlation dimension $D_2$ obtained from
galaxy surveys, showing a general increase with scale. Scaramella et
al.\cite{scara98} analysed a number of subsamples with different
methods, from which we chose one of their largest.  All their results
that include necessary `k-corrections', which account for the effect
of galaxy spectra being redshifted relative to the observer's pass
band, are consistent with $D_2=3$ within their errors.  Also given are
estimates of $D_2$ from the X-Ray Background and the Cosmic Microwave
Background, obtained by normalising a standard CDM model to match
measured anisotropy results (see box). Unlike the other measurements
in this Table, the CMB probes directly the fluctuations in mass.

\begin{table}
\begin{tabular}{llcl}
\hline
                        & Sample      &  $R$ ($h^{-1}$ Mpc)  & $D_2$ \\ 
\hline
Guzzo et al.\cite{gicgh91} & Perseus-Pisces  & 1.0--3.5     & 1.2     \\
                        &                 & 3.5--20      & 2.2     \\
Martinez \& Coles\cite{mc94}& QDOT            & 1.0--10      & 2.25    \\
                        &                 & 10--30       & 2.77    \\
Lemson \& Sanders\cite{ls91}& CfA             & 1.0--30      & 2.0     \\
Mart\'inez et al.\cite{mpmg98}& Stromlo-APM & 30--60       & 2.7--2.9 \\   
Scaramella et al.\cite{scara98}& ESP             & 300--400     & 2.93    \\
\\
X-Ray Background       &                 & $\sim 500$  & $3-D_2=10^{-4}$\\
                                       &&&with $\sigma_8=2$, $\Gamma=0.5$\\
COBE 4 year normalisation &         & $\sim 1000$  & $3-D_2=2\times 10^{-5}$\\
                                       &&&with $\sigma_8=1.4$, $\Gamma=0.5$\\
\hline
\end{tabular}
\end{table}

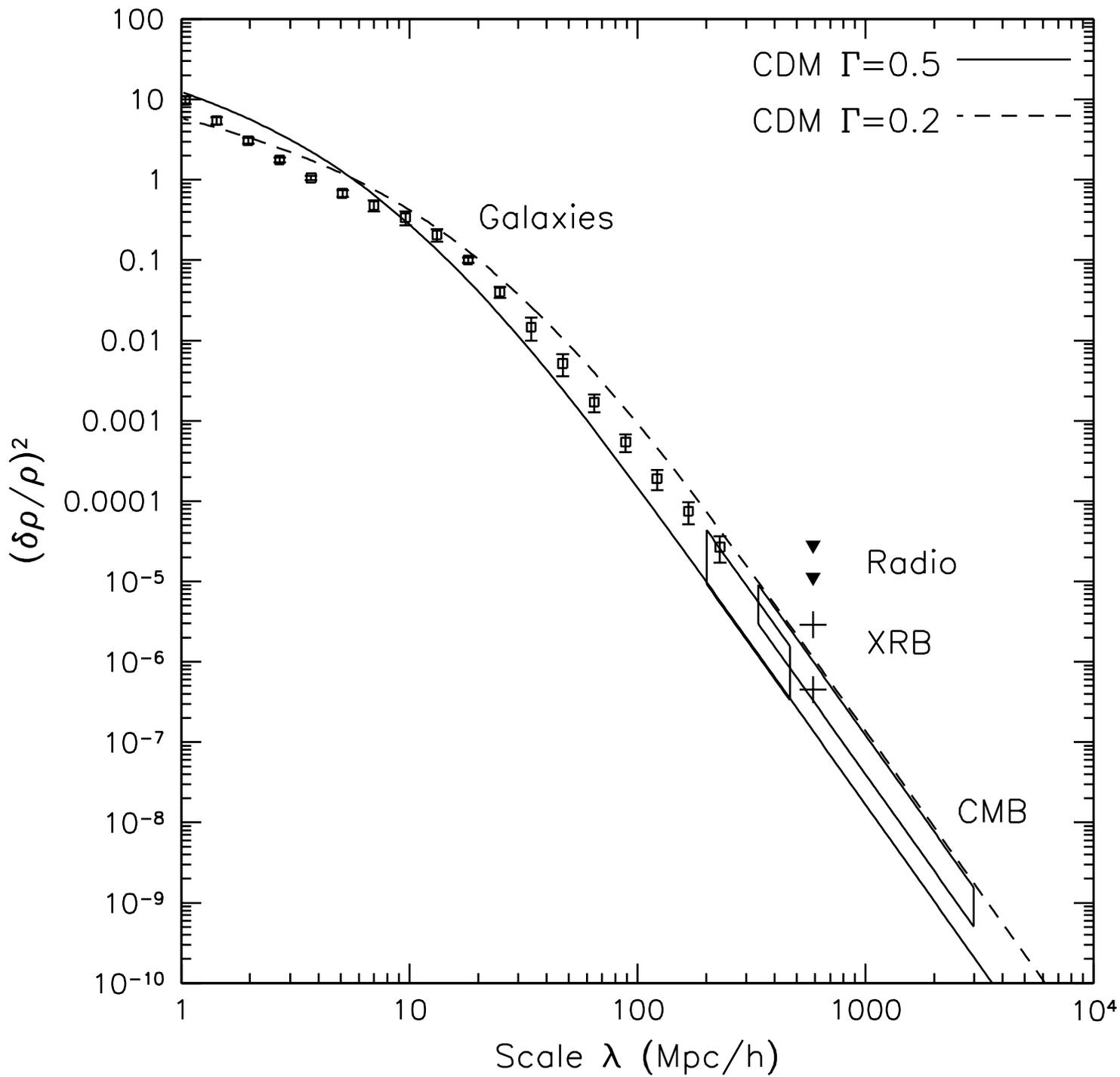
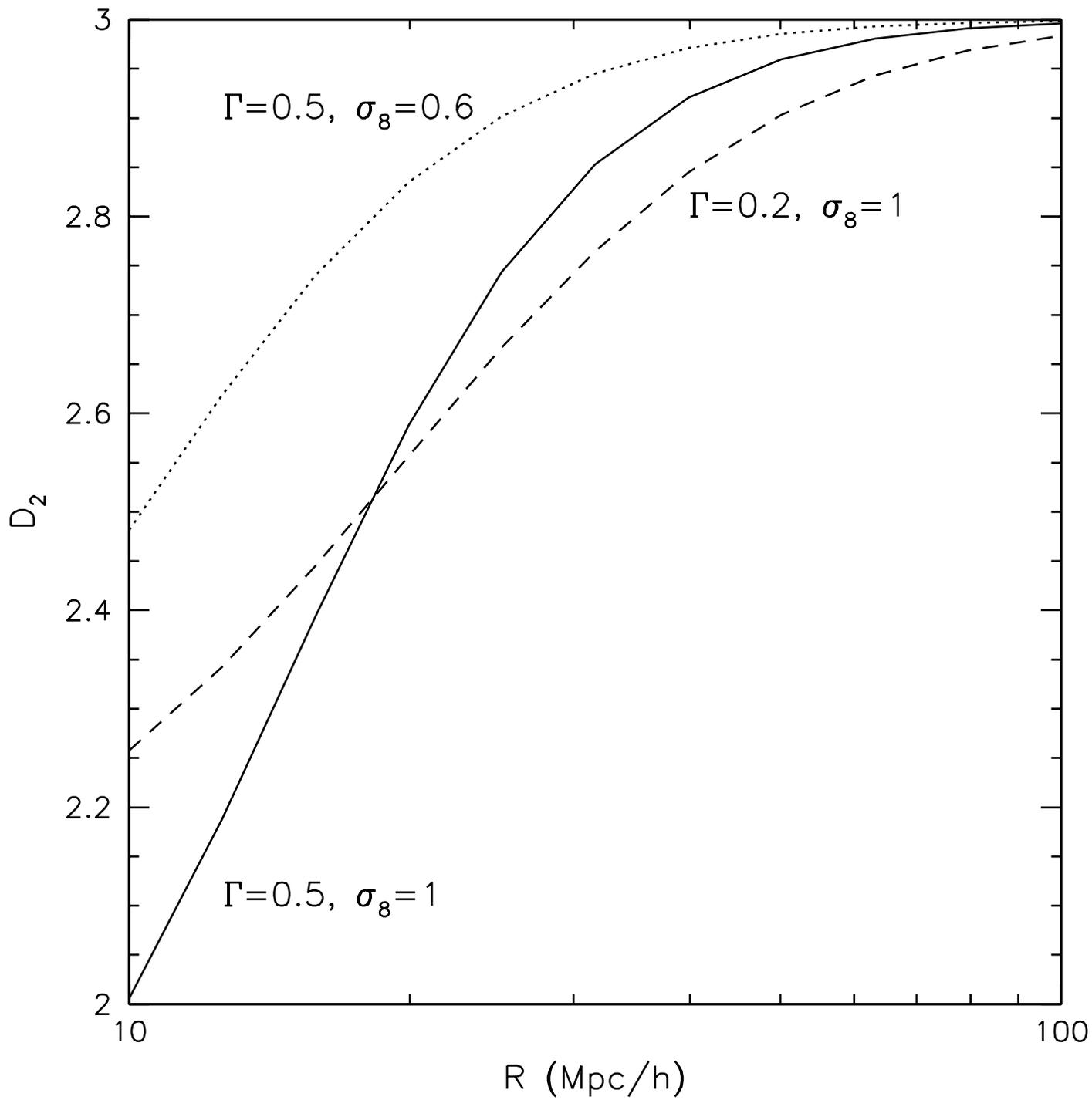
\end{document}